\documentclass[12pt]{iopart}
\usepackage{graphicx}
\begin{document}

\title[Stochastic backgrounds of gravitational waves from extragalactic sources]{Stochastic backgrounds of gravitational waves from extragalactic sources}

\author{R Schneider$^{1}$, S Marassi$^{2}$, V Ferrari$^{2}$}

\address{$^{1}$ INAF, Osservatorio Astrofisico di Arcetri, Largo Enrico Fermi 5, 50125, Firenze, Italy\\
$^{2}$ Dipartimento di Fisica “G.Marconi”, Sapienza Universit`a di Roma 
and Sezione INFN ROMA1, piazzale Aldo Moro 2, I-00185 Roma, Italy}
\ead{raffa@arcetri.astro.it}

\begin{abstract}

%Gravitational waves of cosmological origin could be the result
%of a large variety of astrophysical and cosmological processes
%that develop in the very early Universe. As a consequence,
%our high redshift Universe is expected to be permeated with a
%background of gravitational radiation. Depending
%on their origin, these stochastic gravitational wave
%backgrounds will show different spectral properties
%and features that it is important to investigate in view of a
%possible, future detection. In this contribution, we will review
%recent theoretical predictions for backgrounds produced
%by extragalactic sources and discuss their detectability
%with current and future gravitational wave observatories.

Astrophysical sources emit gravitational waves in a large 
variety of processes occurred since the beginning of star and
galaxy  formation. These waves permeate our high redshift Universe, 
and form a background which is the result of the superposition 
of different components, each associated to a specific astrophysical process.
Each component has different spectral properties
and features that it is important to investigate in view of a
possible, future detection. In this contribution, we will review
recent theoretical predictions for backgrounds produced
by extragalactic sources and discuss their detectability
with current and future gravitational wave observatories.

\end{abstract}

%\pacs{00.00, 20.00, 42.10}
%\submitto{\JPA}

\maketitle

\section{Introduction}

A stochastic background of gravitational waves (GWs) is expected to arise as a super-position 
of a large number of unresolved sources, from different directions in the sky, and 
with different polarizations. It is usually described in terms of the GW spectrum,
\[
\Omega_{\rm GW} = \frac{f}{\rho_{\rm cr}}\frac{d\rho_{\rm GW}}{df}, 
\]
\noindent
where $d\rho_{\rm GW}$ is the energy density of GWs in the frequency range $f$ to $f+df$ and
$\rho_{\rm cr}$ is the critical energy density of the Universe.

Stochastic backgrounds of GWs are interesting to investigate because they (i) represent
a foreground signal for the detection of primordial backgrounds (see the contribution of
M. Maggiore in this conference), and (ii) provide important information on GW sources and
on the properties of distant stellar populations.

The starting point in any calculation of the GW backgrounds produced by extragalactic sources
is the redshift evolution of the comoving star formation rate density. This is generally inferred
from observations which have been recently extended out to redshifts $z \sim 10$ following
the analysis of the deepest near-infrared image ever obtained on the Hubble Ultra Deep Field 
using the new/IR camera on HST (Bouwens et al. 2010).  
The detection of galaxies at $z = 8.4-10$, just 500-600 Myr after recombination and in the heart 
of the reionization epoch, provides clues about galaxy build-up at a time that must 
have been as little as 200-300 Myr since the first substantial star formation (at $z \sim 15-20$). 
The existence of galaxies at $z \sim 10$ shows that galaxy formation was well underway at this  
very early epoch. 

\section{GW backgrounds from extragalactic sources: early predictions}

The most interesting astrophysical sources of gravitational radiation such 
as the core-collapse of a massive star to a black hole, the instabilities of a rapidly 
rotating neutron star or the inspiral and coalescence of a compact binary system are all 
related to the ultimate stages of stellar evolution. Thus, the rate of formation of these 
stellar remnants can be deduced from the rate of formation of the corresponding stellar 
progenitors once a series of assumptions are made regarding the stellar evolutionary track 
followed.

%%%%%%%%%%%%
\begin{figure}
\includegraphics[width=8.5cm,angle=360]{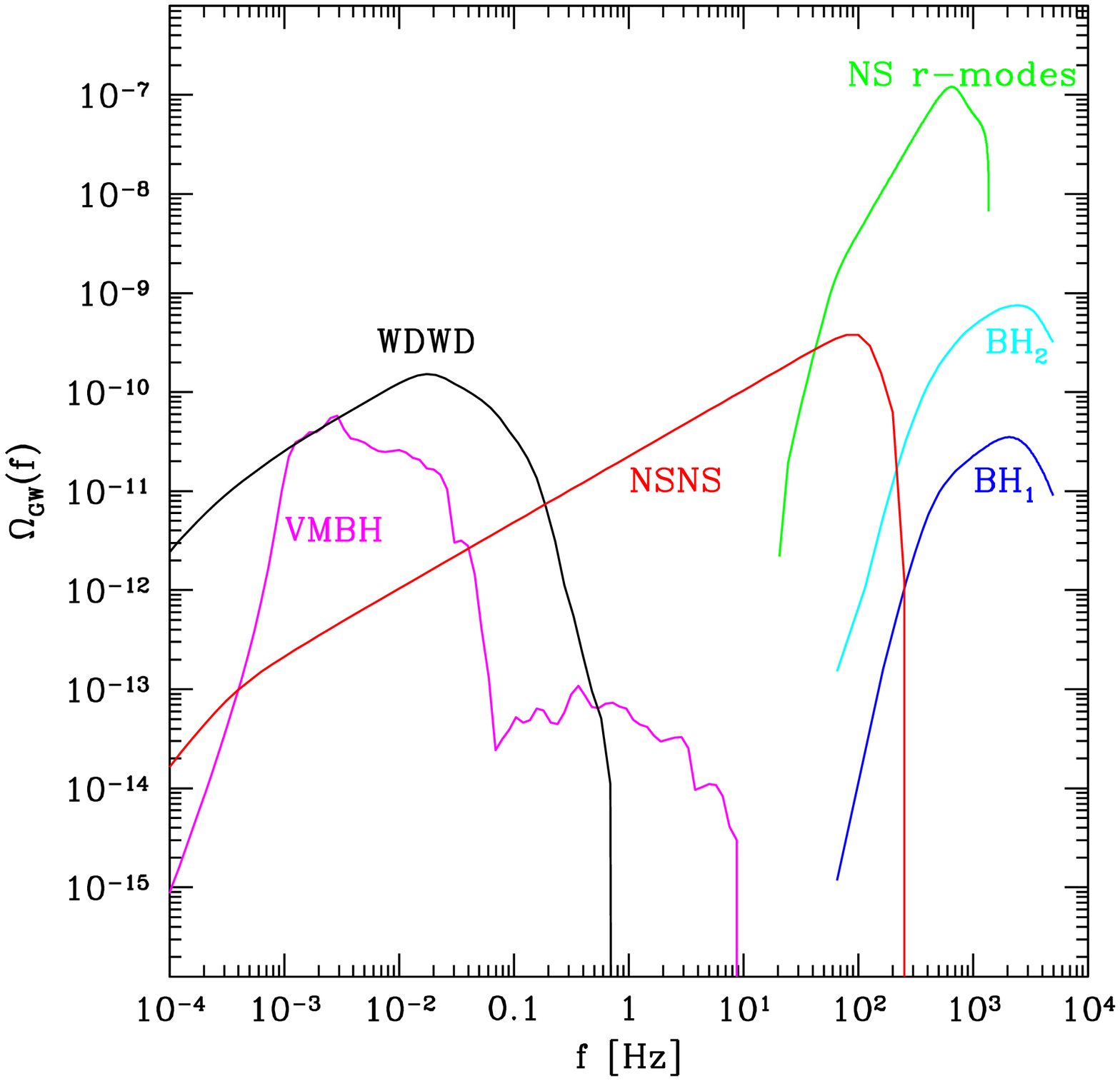}
\includegraphics[width=8.5cm,angle=360]{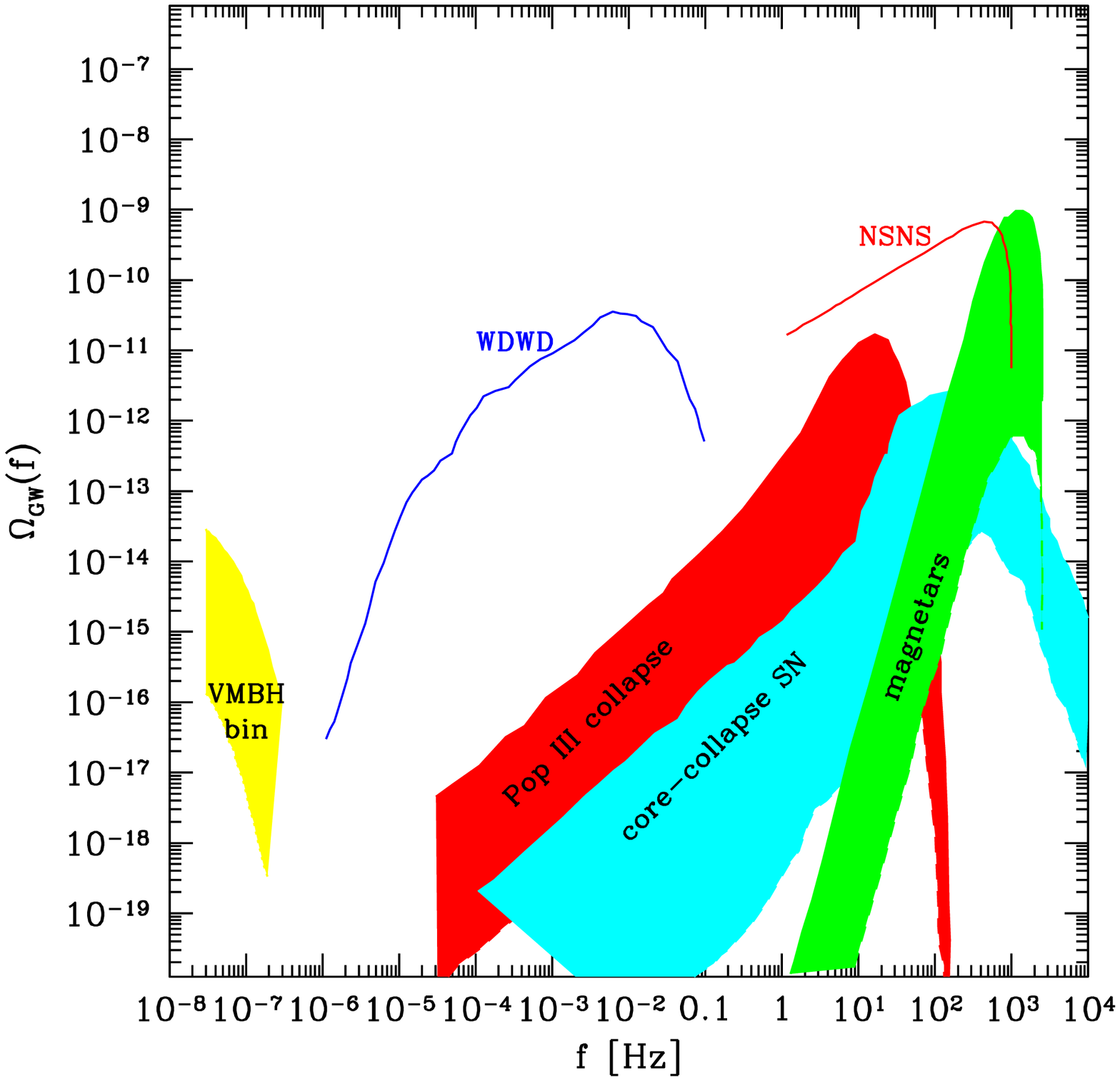}
\caption{{\it Left panel:} collection of GW background signals for
different sources, such as double neutron stars (NSNS) and white dwarfs (WDWD, Schneider et al. 2001),
rapidly rotating neutron stars (NS r-modes, Ferrari et al. 1999b), black hole collapse assuming two 
possible values for the specific angular momentum (BH1, BH2, Ferrari et al. 1999a), collapse of 
super-massive stars to black hole (VMBH, Schneider et al. 2000). {\it Right panel:} collection of GW
backgrounds predicted by additional studies for double neutron stars (NSNS, Regimbau \& Chauvinaeu 2007),
double white dwarfs (WDWD, Farmer \& Phinney 2003), collapse of Population II (core-collapse SN) and 
III stars (Pop III collapse, Buonanno et al. 2005), magnetars (Regimbau \& de Freitas
Pacheco 2006), binary very massive black hole systems (VMBH bin, Sesana et al. 2008).}
\label{fig:GWback}
\end{figure}
%%%%%%%%%%%%

A few years ago we have published estimates of the stochastic backgrounds produced by 
astrophysical sources, concentrating in particular on the (i) gravitational collapse 
of a massive star that leads to the formation of a black hole (Ferrari, Matarrese \& Schneider 1999a), 
(ii) the emission of GWs by newly born rapidly rotating neutron stars (Ferrari, Matarrese \& Schneider 1999b),
(iii) gravitational collapse of a super-massive star to black hole (Schneider et al. 2000), 
(iv) the orbital decay of various degenerate binary systems that loose angular momentum as a 
consequence of the emission of gravitational radiation (Schneider et al. 2001).

While to compute the first two classes of backgrounds we could simply derive the black holes and 
neutron stars formation rate as a 
function of redshift from the corresponding star formation rate (estimated from observational data) via an assumed stellar initial mass function (IMF), 
the last two classes required additional information from theoretical models for the evolution of primordial objects (VMBHs), and from a 
binary population synthesis model (compact binaries). We refer the interested reader to the original papers for further details.
The resulting GW backgrounds are presented in the left panel of Fig.~\ref{fig:GWback}. In the right panel of the same figure, we
also show a collection of signals predicted by additional studies on double compact stellar binaries 
(Farmer \& Phinney 2003; Regimbau \& Chauvineau 2007), on the collapse of Population II and III stars (Buonanno et al. 2005), on 
strongly magnetized neutron stars (Regimbau \& de Freitas Pacheco 2006), and on very massive black hole binary systems (Sesana, 
Vecchio \& Colacino 2008). Shaded regions illustrate the sensitivity of the corresponding signal to variations of critical parameters.

\section{GW backgrounds from Population III and II stars}

Our present understanding of Pop~III star formation suggests that these
stars are not necessarily confined to form in the first dark
matter halos at redshift $z > 20$ but may continue to form during
cosmic evolution in regions of sufficiently low metallicity, with
$Z < Z_{\rm cr} = 10^{-5 \pm 1}  Z_{\odot}$.  
At these low metallicities, the reduced cooling efficiency and large
accretion rates favor the formation of massive stars, with characteristic 
masses $\ge 100 M_{\odot}$ (Bromm et al. 2001; Schneider et al. 2002, 2006a;
Omukai et al. 2005). With the exception of the dynamical range $(140 - 260) M_{\odot}$,
where they are expected to explode as pair-instability SNe (Heger \& Woosley 2002), 
these stars are predicted to collapse to black holes of comparable masses and thus
to be efficient sources of GWs. Earlier estimates of the extragalactic background
generated by the collapse of the first stars, discussed in the previous section,
were limited by our poor understanding of their characteristic masses and formation
rates. In particular, the Pop~III star formation rate and the
cosmic transition between Pop~III and Pop~II stars is regulated by the rate
at which metals are formed and mixed in the gas surrounding the first
star forming regions, a mechanism that we generally refer to as chemical feedback.

Semi-analytic studies which implement chemical feedback generally find that, due
to inhomogeneous metal enrichment, the transition is extended in time, with coeval
epochs of Pop~III and Pop~II star formation, and that Pop~III stars can continue to
form down to moderate redshifts, $z < 5$ (Scannapieco, Schneider \& Ferrara 2003;
Schneider et al. 2006b). To better assess the validity of these semi-analytic models, 
Tornatore, Ferrara \& Schneider (2007)
have performed a set of cosmological hydrodynamic simulations with an improved
treatment of chemical enrichment. In the simulation, it is
possible to assign a metallicity-dependent stellar IMF. When
$Z>Z_{\rm cr}$ Pop~II stars are assumed to form according to a Salpeter IMF
$\Phi(M)\propto M^{-(1+x)}$ with $x=1.35$ and lower
(upper) mass limit of $0.1 M_{\odot}$ ($100 M_{\odot}$). Stars with masses in the $8-40 M_{\odot}$
range explode as Type-II SNe and the explosion energy and metallicity-dependent metal yields
are taken from Woosley \& Weaver (1985). Stars with masses $>40 M_{\odot}$ do not contribute
to metal enrichment as they are assumed to directly collapse to black holes.
Very massive Pop~III stars form in regions where $Z < Z_{\rm cr}$; since theoretical models
do not yet provide any indication on the shape of the Pop~III IMF, Tornatore et al. (2007)
adopt a Salpeter IMF shifted to the mass range $100 - 500 M_{\odot}$; only stars in the
pair-instability range ($140 - 260 M_{\odot}$) contribute to metal-enrichment and the
metal yields and explosion energies are taken from Heger \& Woosley (2002).
The simulation allows to follow metal enrichment properly accounting for the finite
stellar lifetimes of stars of different masses, the change of the stellar IMF and
metal yields. The numerical schemes adopted to simulate metal transport ad diffusion
are discussed in Tornatore et al. (2007) to which we refer the interested reader
for more details.

\begin{figure}
\includegraphics[width=8.5cm,angle=360]{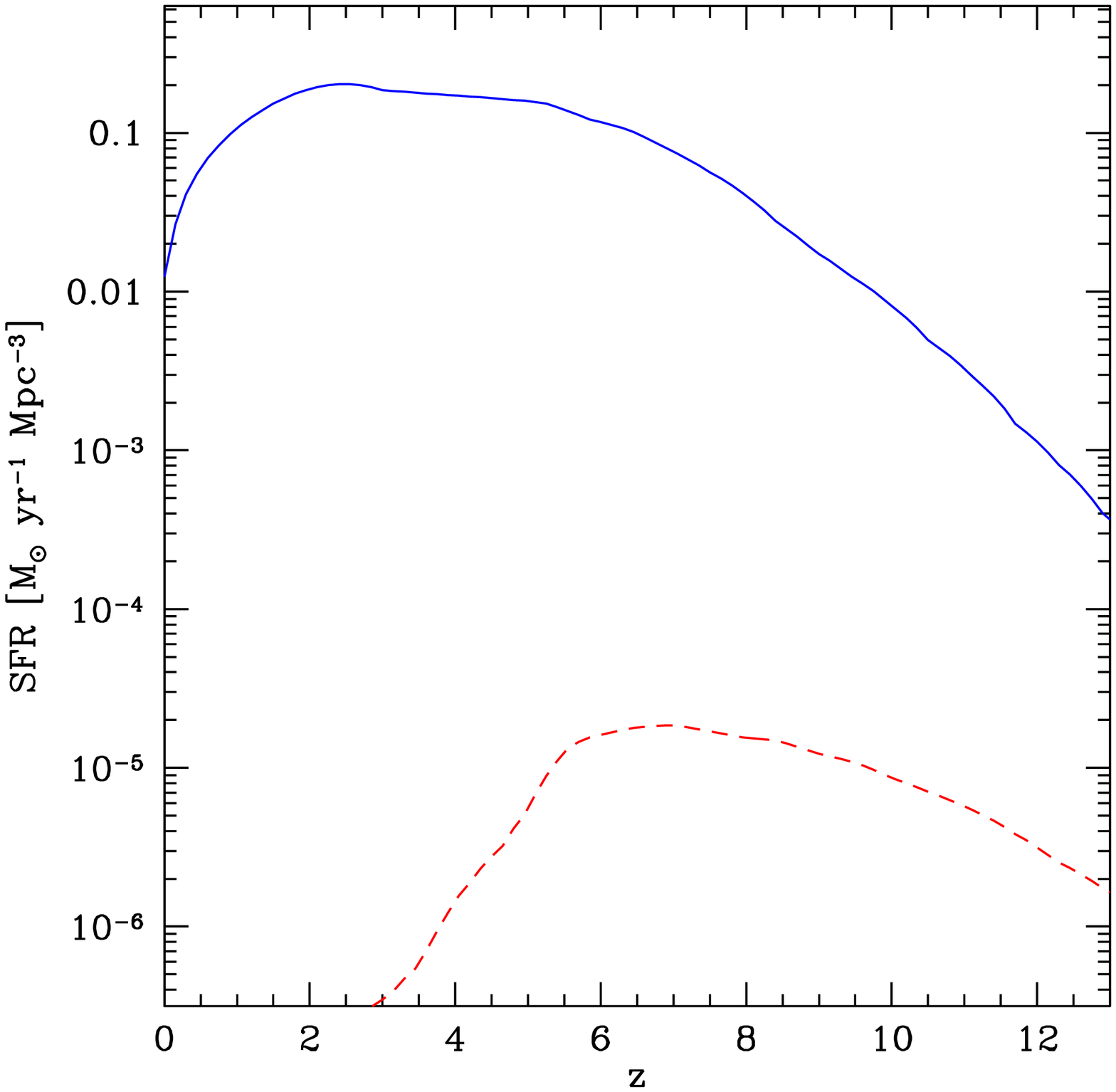}
\includegraphics[width=8.5cm,angle=360]{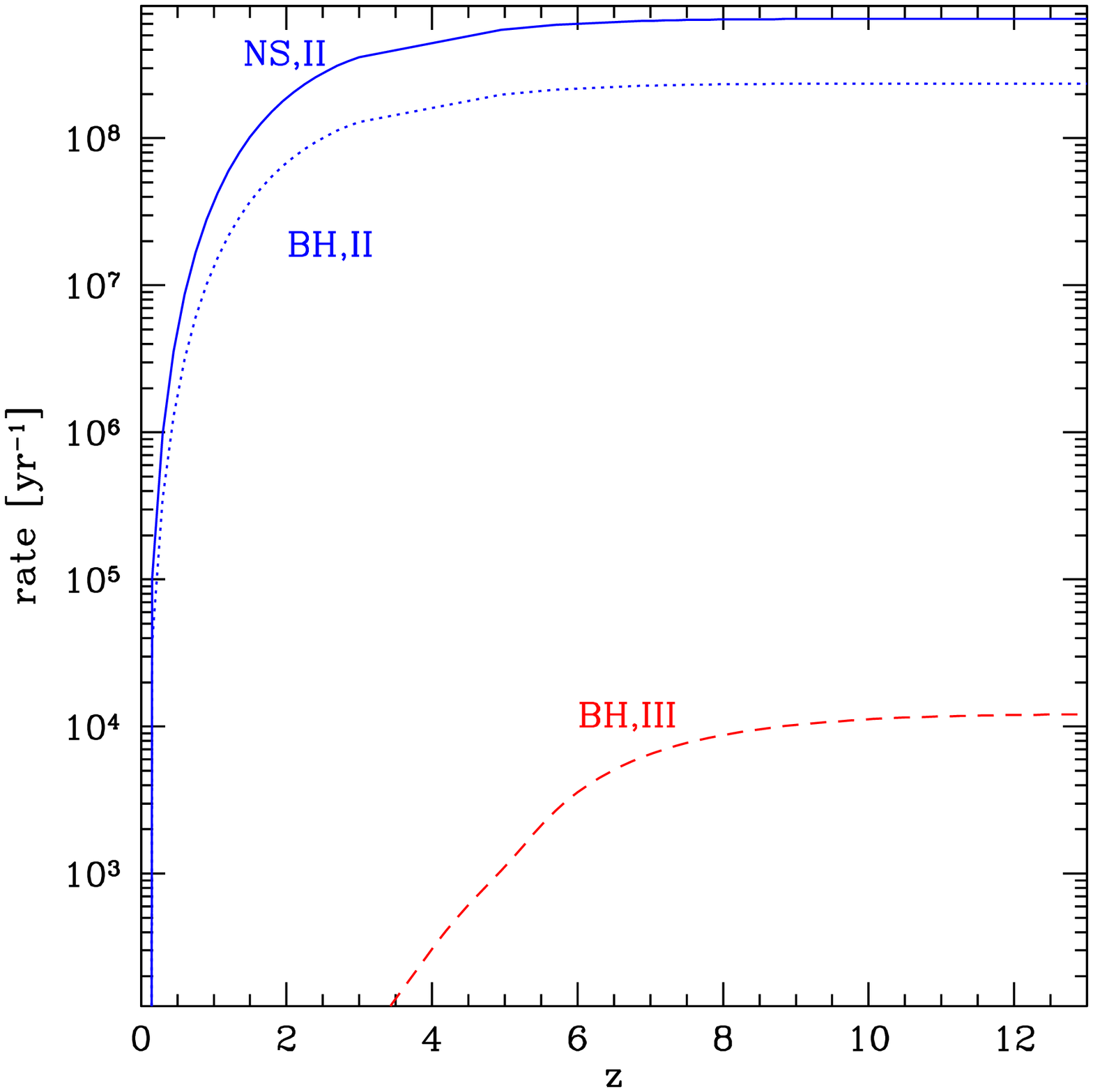}
\caption{
{\it Left panel:} redshift evolution of the comoving star formation rate density (solid
lines). The dotted line shows the contribution of Pop~III stars (see text). 
{\it Right panel:} redshift evolution of the number of gravitational wave sources formed per unit time within a
comoving volume. The top curves show the contribution of Pop~II stars leaving behind neutron
stars (NS,II solid line) and black holes (BH,II dotted); the bottom dotted line represents
the contribution of Pop~III stars collapsing to black holes (BH,III dashed). 
}
\label{fig:sfr}
\end{figure}

The top panel in Fig.~\ref{fig:sfr} shows the redshift evolution of the cosmic star formation rate
and the contribution of Pop~III stars predicted by the simulation\footnote{The results
shown in Fig.~\ref{fig:sfr} refer to the fiducial run in Tornatore et al. (2007) with
a box of comoving size $L=10 h^{-1}$~Mpc and $N_{\rm p} = 2 \times 256^3$ (dark+baryonic)
particles.}. In this model, the critical metallicity which defines the Pop~III/Pop~II
transition is taken to be $Z_{\rm cr} = 10^{-4} Z_{\odot}$, at the upper limit of the
allowed range. However, additional runs show that decreasing the critical metallicity
to $10^{-6} Z_{\odot}$ reduces the Pop~III star formation rate by a factor $< 10$.
It is clear from the figure that Pop~II stars always dominate the cosmic star formation
rate; however, in agreement with previous semi-analytic studies, the simulation shows
that Pop~III stars continue to form down to $z < 5$, although with a decreasing
rate. Over cosmic history, the fraction of baryons processed by Pop~III stars is predicted
to be $f_{b} = 2 \times 10^{-6}$.

%%%%%%%%%%%%
%%%%%%%%%%%%

\subsection{GW emission from $[8 - 20] M_{\odot}$ Pop II stars} 

\begin{figure}
\includegraphics[width=8.0cm,angle=360]{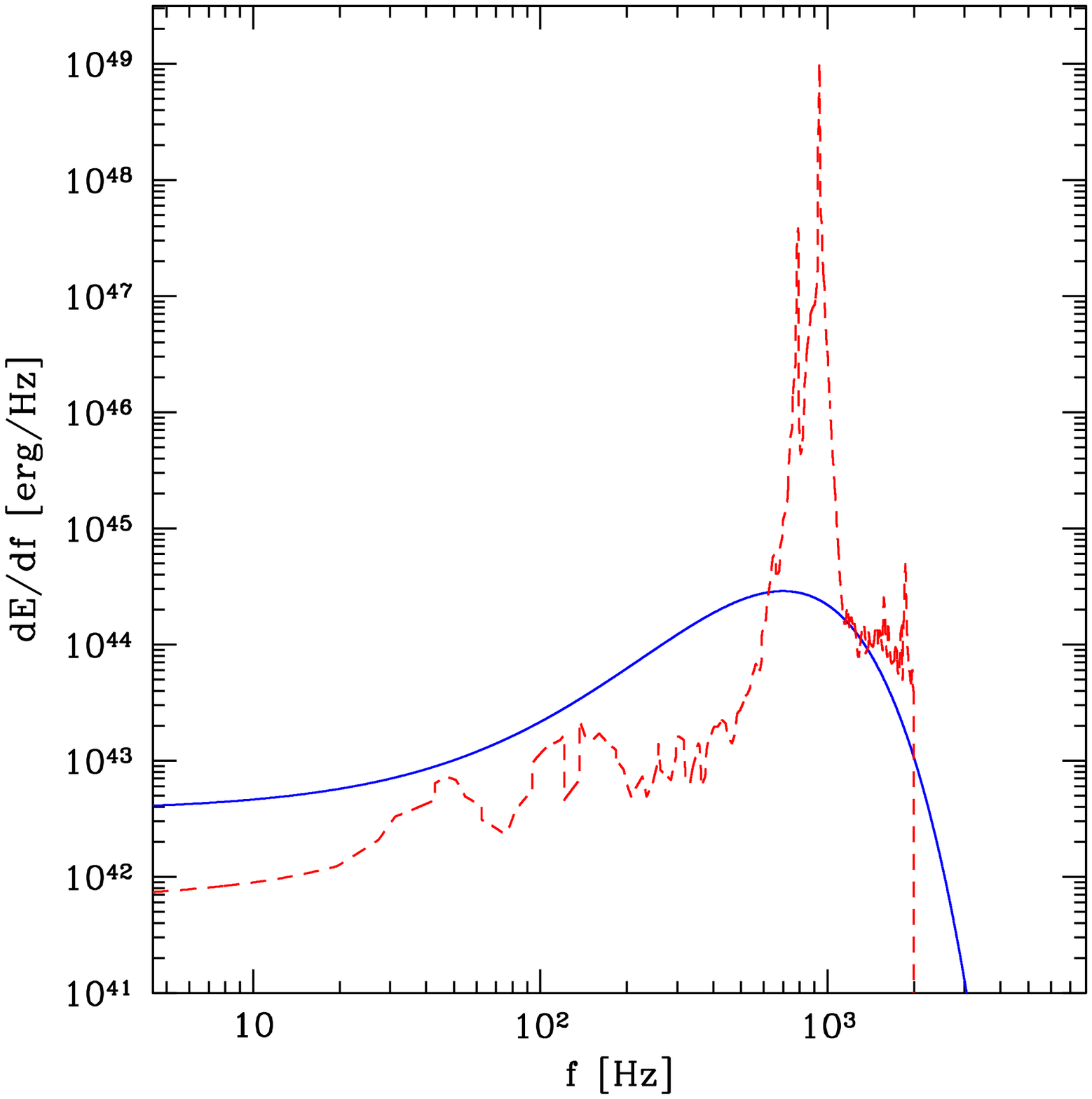}
\includegraphics[width=8.0cm,angle=360]{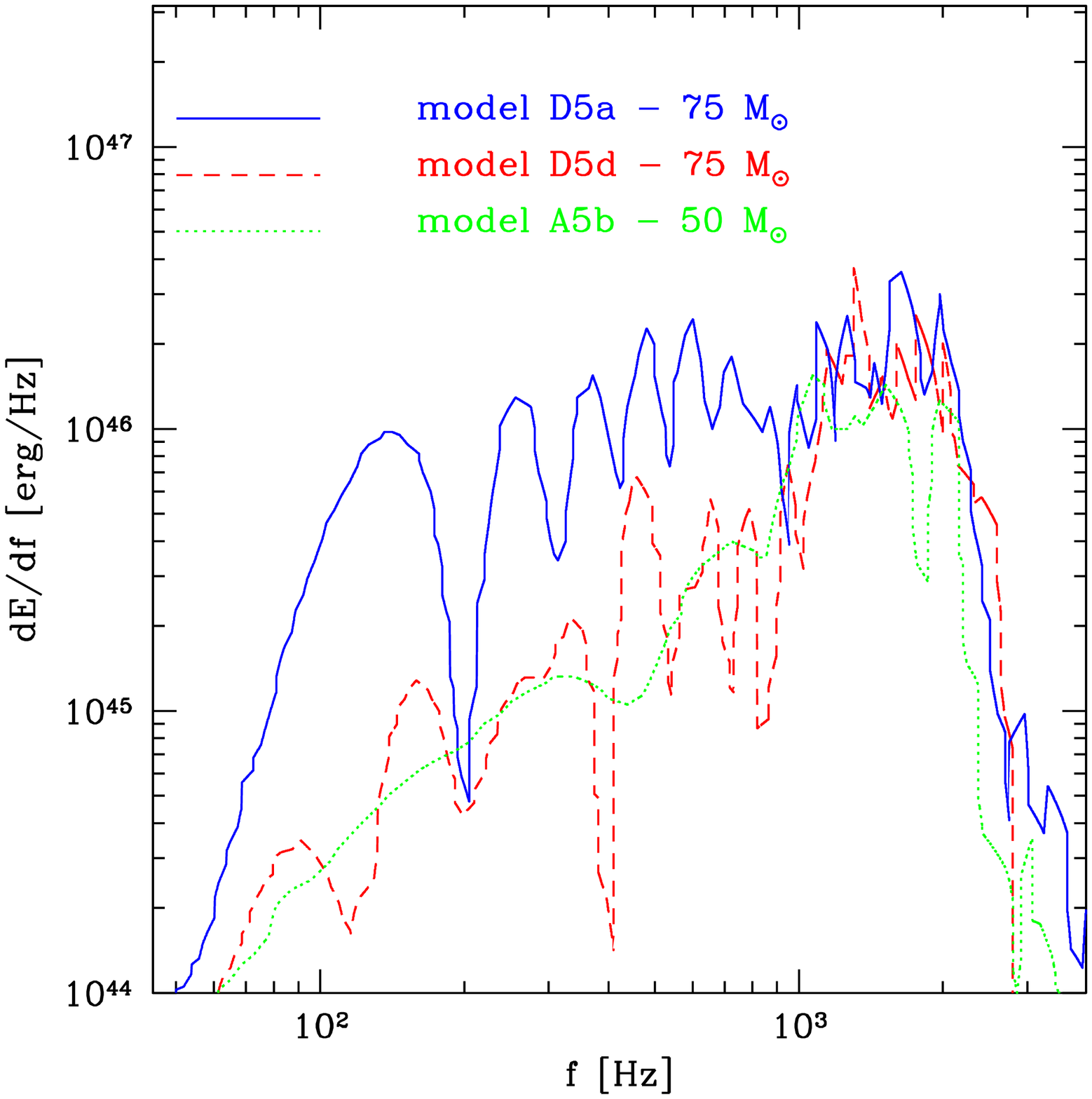}
\includegraphics[width=8.0cm,angle=360]{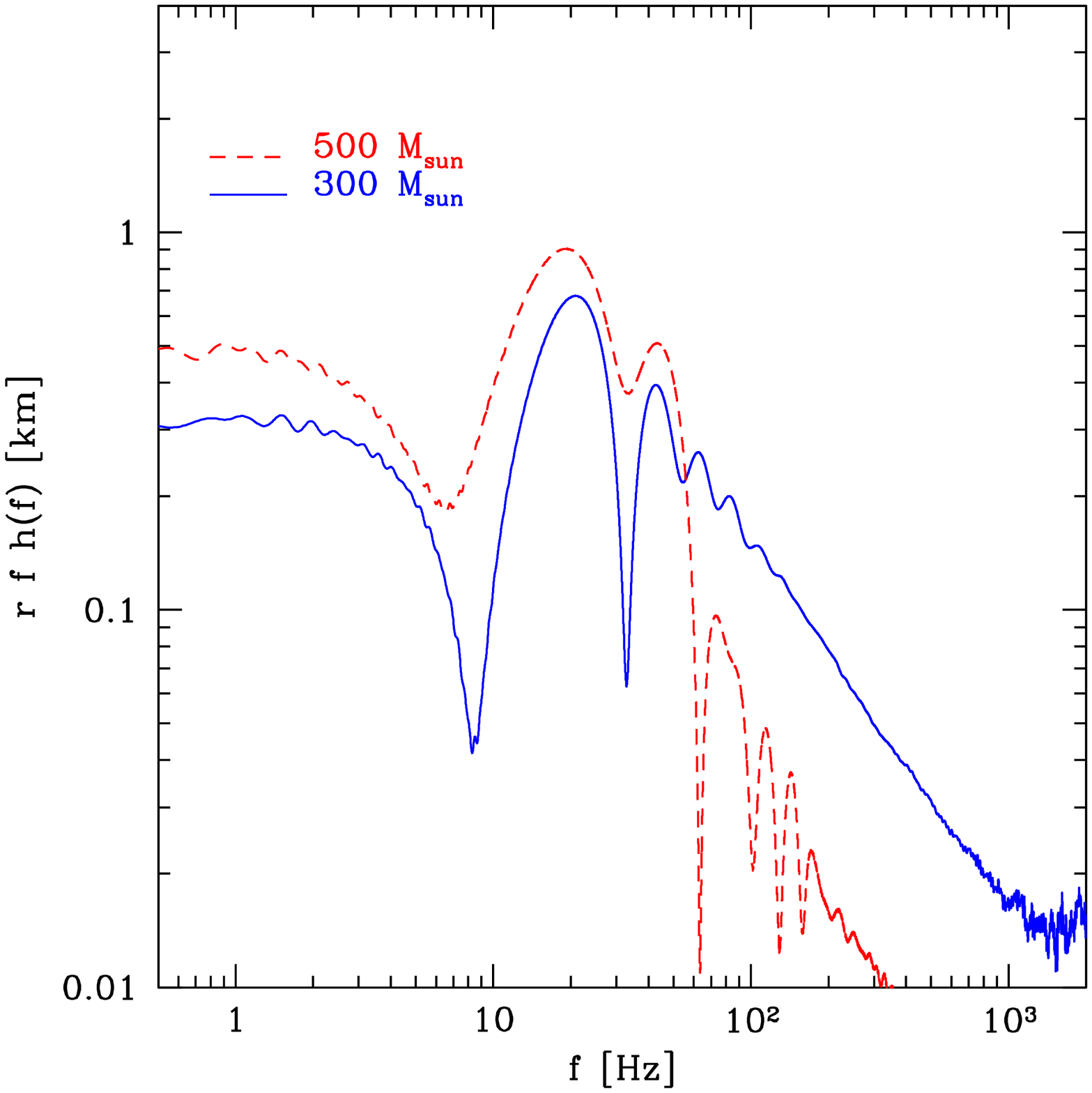}
\caption{
{\it Upper left panel}: template GW energy spectrum for core collapse to neutron stars. The solid 
line is the rotating 15 $M_{\odot}$ progenitor model of Muller et al. (2004), the dashed line
is the g-mode spectrum for a 25 $M_{\odot}$ progenitor model of Ott et al. (2006). 
{\it Upper right panel:} template GW energy spectrum for core collapse to black holes. Three selected
models from the analysis of Sekiguchi \& Shibata (2005) are shown, which differ for the initial
progenitor mass and equation of state. The resulting dynamical evolution and GW emission is 
different (see text).
{\it Lower panel:} GW spectral distribution taken from Suwa et al. (2007) for two different 
template progenitor models with $300$ and $500 M_{\odot}$.    
}
\label{fig:gwsingle}
\end{figure}

The first class of sources we have considered are core-collapse SNe leaving 
behind a NS remnant. To model the GW emission from these sources we have
used the waveforms predicted by Muller et al. (2004) using sophisticated 
2D numerical simulations with a detailed implementation of neutrino transport and
neutrino-matter interactions.  In particular, we adopt as our template spectrum 
that of a rotating 15 $M_{\odot}$ star which also gives the most optimistic GW signals,
which corresponds to a GW emission efficiency of $1.8 \times 10^{-8} M_{\odot} c^2$. 
The waveform is shown in the upper left panel of Fig.~\ref{fig:gwsingle}.

%%%%%%%%%%%%
%%%%%%%%%%%%

As an alternative upper limit to the GW emission associated to these sources 
we have also assumed as a template for our single source spectrum the waveform
obtained by Ott et al. (2006); in this scenario, based
on axisymmetric Newtonian core-collapse supernova simulations,  
the dominant contribution to gravitational wave emission  is due to the
oscillations of the protoneutron star core. These  are predominantly
g-modes oscillations, which are excited  hundreds of milliseconds after bounce 
by turbulence and by accretion downstreams from the SN shock
(undergoing the Standing-Accretion-Shock-Instability, SASI),
and typically last for several hundred milliseconds.
In this phase a strong GW emission has been shown to occur,
with a corresponding energy ranging  from 
$1.5 \times 10^{-8} M_{\odot} c^2$ for a 11 $M_{\odot}$ progenitor, to 
$8.2 \times 10^{-5} M_{\odot} c^2$ for a 25 $M_{\odot}$ progenitor. In the latter case, 
the largest emission is caused by the more massive iron core, higher post-bounce 
accretion rates and higher pulsation frequencies and amplitudes. 
In these models the contribution of the anisotropic neutrinos emission is found to be 
negligible.
If we take the 25 $M_{\odot}$ progenitor model, we can assume this to be an 
upper limit to the GW emission from these sources. 
The corresponding waveform is also shown in the upper left panel of Fig.~\ref{fig:gwsingle}.

\subsection{GW emission from $[20 - 100] M_{\odot}$ Pop II stars} 

Progenitors with masses between 20 and 100 $M_{\odot}$ may lead to a prompt collapse 
to BH or to the formation of a proto-neutron star that, due to signficant fall-back,
will end its life as a black hole. To model the GW emission from these events we have 
used the results of numerical simulations performed in full GR by 
Sekiguchi \& Shibata (2005) of a grid of rotating stellar cores resulting from the 
evolution of stellar progenitors with masses 50-100 $M_{\odot}$.  We have selected 
three representative models: D5a and D5d are associated to 75 $M_{\odot}$ stars, A5b 
to a 50 $M_{\odot}$ star. These models, which are shown in the upper right panel of 
Fig.~\ref{fig:gwsingle}, also differ for the equations of state, which 
translate into distinct collapse dynamics.  The gravitational energy emitted 
is typically in the range $[2 - 3] \times 10^{-7} M_{\odot} c^2$.

\subsection{GW emission from $[100 - 140] M_{\odot}$ and $[260 - 500] M_{\odot}$ Pop III stars} 

The collapse of Pop III stars to black holes has been suggested to be a much more 
efficient source of GWs than todays’ SN population: according to 2D simulations 
of a zero-metallicity 300 $M_{\odot}$ stars which take the effects of GR and 
neutrino transport into account, the estimated total energy released in GWs ranges between 
$[2 \times 10^{-4} - 2 \times 10^{-3}] M_{\odot} c^2$ (Suwa et al. 2007, Fryer et al. 2001), 
with a factor of ten discrepancy among the two analyses ascribed to different initial 
angular momentum distribution. Since we want to be conservatives, we take the gravitational 
waveforms predicted by the most recent analysis of Suwa et al. (2007) which we show in the 
lower panel of Fig.~\ref{fig:gwsingle}.

\section{Results}

\begin{figure}
\includegraphics[width=8.5cm,angle=360]{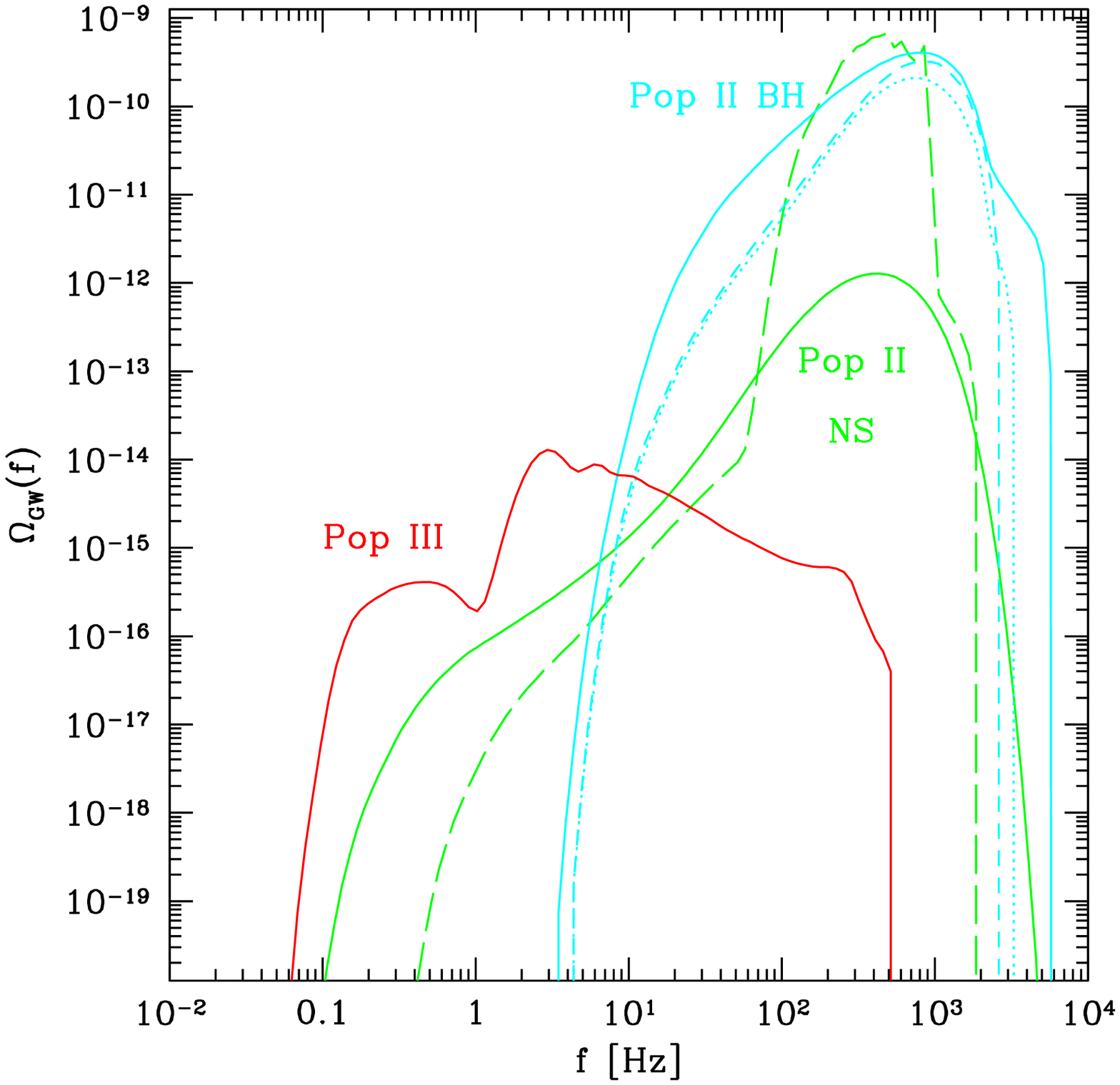}
\includegraphics[width=8.5cm,angle=360]{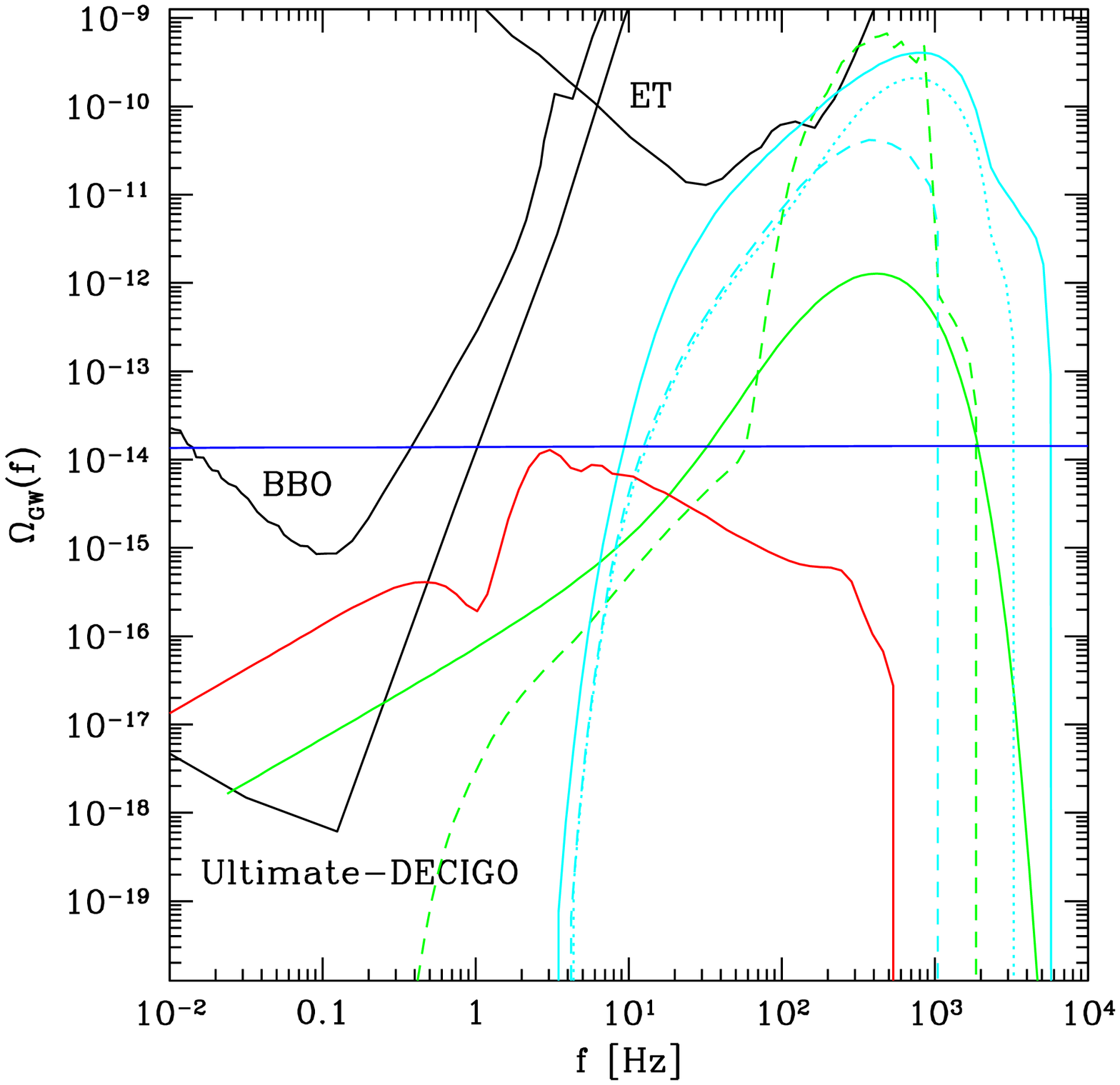}
\caption{
{\it Left panel}: the function $\Omega_{\rm GW}$ as a function of frequency for stochastic backgrounds associated to 
Pop~III/II stars (see text).
{\it Right panel}: same as in the left panel but applying the zero-frequency limit to the background signals where
the gravitational wave emission below $1~Hz$ is dominated by the neutrino contribution. Superimposed on the astrophysical
signals are the sensitivity curves for the next generation gravitational telescopes (BBO, Ultimate-DECIGO, ET) and the
upper limit on the stochastic background generated during the inflationary epoch (see text).} 
\label{fig:gwback_new}
\end{figure}

In Fig.~\ref{fig:gwback_new}, we show the function $\Omega_{\rm GW}$ evaluated
integrating the single source spectra described above over the corresponding rate evolution 
(Marassi, Schneider \& Ferrari 2009),

\[
\Omega_{\rm GW}(f)=\frac{f}{c^3\rho_{cr}}\left[\frac{dE}{dS df dt}\right] ,
\]

\noindent
where $dE/(dS df dt)$ is the spectral energy density
of the stochastic background,
\[
\frac{dE}{dS df dt}=\int^{z_f}_{0} \int^{M_f}_{M_i} dR(M,z)
 \big{<}\frac{dE}{dS df}\big{>} ,
\]

\noindent
$dR(M,z)$ is the differential source formation rate,
\[
dR(M,z)=\frac{\dot{\rho}_\star(z)}{(1+z)}\frac{dV}{dz}\Phi(M)dMdz,
\]
\noindent
and $\big{<}\frac{dE}{dS df}\big{>}$ is the locally measured average
energy flux emitted by a source at distance $r$. For
sources at redshift $z$ it becomes,
\[
\big{<}\frac{dE}{dS df}\big{>}=\frac{(1+z)^2}{4\pi d_L(z)^2}\frac{dE^{e}_{GW}}{df_e}[f(1+z)]
\]
\noindent
where $f=f_e(1+z)^{-1}$ is the redshifted emission frequency $f_e$,
and $d_L(z)$ is the luminosity distance to the source.
The redshift evolution of the GW source formation rates, integrated
over the comoving volume element, are shown in the right panel of Fig.~\ref{fig:sfr}.

We see that below $\sim 10$~Hz the Pop~III GWB dominates over
the Pop II background; its  maximum
amplitude is $\Omega_{\rm GW} \simeq 10^{-14}$ at $f=2.74$~Hz.
At larger frequencies the background produced by  Pop~II stars is
much larger than that of  Pop~III.
Stars with progenitors in the range (20-100)$M_{\odot}$ contribute with  a nearly
monotonically increasing behavior, reaching amplitudes
$\Omega_{\rm GW} \simeq 4\times 10^{-10}$ at $f\in (759-850)$~Hz.
For stars with progenitors in the range (8-25)$M_{\odot}$ the GWB
has a shape which depends on the waveform used as representative of the
collapse: that obtained using the waveform of Muller et al. (2004)
is quite smooth and peaks at $f = 387$~Hz, with amplitude $\Omega_{\rm GW} \simeq 10^{-12}$.
The background obtained using  waveforms produced by the  core-collapse model
of Ott et al. (2006), instead, is comparable in amplitude with those produced by more massive
progenitors, reaching  the maximum amplitude $\Omega_{\rm GW} \simeq 7\times 10^{-10}$ at $f= 485$ Hz.

%%%%%%%%%%%%
%%%%%%%%%%%%

In the right panel of Fig.~\ref{fig:gwback_new}, we have applied to the same signals the so-called
zero-frequency limit: in fact, current numerical
simulations describe the source collapse at most for a few seconds, and
cannot predict the emission below a fraction of Hz. To evaluate the background in the low frequency region
$f \leq$ 0.1 Hz, it is customary to extend the single source waveform $f|\tilde{h}(f)|$ to lower
frequencies, where the emission is dominated by the neutrino signal\footnote{This is not done for the 
single source spectra derived by Sekiguchi \& Shibata (2005) and Ott et al. (2006), as in these 
computations neutrinos contribution is neglected or subdominant and the zero-frequency limit would not be appropriate.}.
Superimposed on the predicted signals, we also show the foreseen sensitivity curves of the next generation
of GW telescopes (BBO, Seto 2006; Ultimate-DECIGO, Kanda, private communication;  ET, Regimbau, private communication). 
All the sensitivity curves are obtained assuming correlated analysis
of the outputs of independent spacecrafts (or of different ground-based
detectors). 
It is clear for the figure that BBO has no chance
to detect the background signals produced by Pop~III/Pop~II stars; these signals, 
however, are within the detection range of Ultimate-DECIGO and are marginally detectable by ET. 

The horizontal line in the same panel shows the upper limit on the stochastic background generated during the inflationary epoch 
taken from Abbott et al. (2009). Thus, the Pop~III/II signals dominate over the primordial background
only for frequencies $\ge 3$~Hz. Even below this frequency range, however, it may be possible to discriminate the astrophysical backgrounds 
from dominant instrumental noise or from primordial signals exploiting the non-Gaussian (shot-noise) structure of the former,
which are characterized by duty-cycles of the order of $10^{-2}$ ($10^{-4}$) for Pop~II (Pop~III) sources (see Marassi et al. 2009). 

In particular Seto (2008) has proposed
a method to check if the detected GWB is
an inflation-type background or if it is contaminated
by undetectable weak burst signals from Pop~III/Pop~II
collapse. If, in the future, it will be possible
to extract information on the Pop~III GWB
component, we can have a tool to investigate
the formation history of these first stars.

\section*{Acknowledgments}
Stefania Marassi thanks the Italian Space Agency (ASI) for the support. This work is funded 
with the ASI CONTRACT I/016/07/0.

\end{document}